\documentclass[%
 aps, pra, reprint,
 superscriptaddress, 
 showpacs, nopreprintnumbers,
 longbibliography,
 amsmath, amssymb, floatfix
]{revtex4-2}

\usepackage[T1]{fontenc}
\usepackage{graphicx}
\usepackage{dcolumn}
\usepackage{bm}
\usepackage{hyperref}
\usepackage{physics}
\usepackage[usenames,dvipsnames]{xcolor}
\usepackage{booktabs}

\DeclareMathOperator{\psim}{\psi_\mathrm{C}}
\DeclareMathOperator{\psil}{\psi_\mathrm{ED}}
\DeclareMathOperator{\psila}{\psi_{\mathrm{ED}_1}}
\DeclareMathOperator{\psilb}{\psi_{\mathrm{ED}_2}}
\DeclareMathOperator{\psimtol}{\psi_{\mathrm{C} \rightarrow \mathrm{ED}}}
\DeclareMathOperator{\psimtola}{\psi_{\mathrm{C} \rightarrow \mathrm{ED_1}}}
\DeclareMathOperator{\psimtolb}{\psi_{\mathrm{C} \rightarrow \mathrm{ED_2}}}

\newcommand{\mytitle}{Accuracy of quantum simulators with ultracold dipolar molecules: \\ a quantitative comparison between continuum and lattice descriptions}

\begin{document}

\title{\mytitle}

 \author{Michael Hughes}
  \email{michael.hughes@physics.ox.ac.uk}
  \affiliation{Clarendon Laboratory, University of Oxford, Parks Rd, Oxford OX1 3PU, United Kingdom.}

 \author{Axel U. J. Lode}
\affiliation{Institute of Physics, Albert-Ludwig University of Freiburg, Hermann-Herder-Strasse 3, 79104 Freiburg, Germany.}

\author{Dieter Jaksch} 
\affiliation{Institut f\"{u}r Laserphysik, Universit\"{a}t Hamburg, 22761 Hamburg, Germany.}
\affiliation{The Hamburg Centre for Ultrafast Imaging, Hamburg, Germany.}
\affiliation{Clarendon Laboratory, University of Oxford, Parks Rd, Oxford OX1 3PU, United Kingdom.}

\author{Paolo Molignini}
  \email{paolo.molignini@fysik.su.se}
\affiliation{T.C.M. group, Cavendish Laboratory, University of Cambridge, Cambridge CB3 0HE, United Kingdom.}
\affiliation{Department of Physics, Stockholm University, AlbaNova University Center, 106 91 Stockholm, Sweden}

\date{\today}

\begin{abstract}
With rapid progress in control and manipulation of ultracold magnetic atoms and dipolar molecules, the quantum simulation of lattice models with strongly interacting dipole-dipole interactions (DDI) and high densities is now within experimental reach. 
This rapid development raises the issue about the validity of quantum simulation in such regimes.
In this study, we address this question by performing a full \emph{quantitative} comparison between the continuum description of a one-dimensional gas of dipolar bosons in an optical lattice, and the single-band Bose-Hubbard lattice model that it quantum simulates.
By comparing energies and density distributions, and by calculating direct overlaps between the continuum and lattice many-body wavefunctions, we demonstrate that in regimes of strong DDI and high densities the continuum system fails to recreate the desired lattice model.
Two-band Hubbard models become necessary to reduce the discrepancy observed between continuum and lattice descriptions, but appreciable deviations in the density profile still remain.
Our study elucidates the role of strong DDI in generating physics beyond lowest-band descriptions and should offer a guideline for the calibration of near-term dipolar quantum simulators.

\end{abstract}
\maketitle

\section{Introduction}
Atomic and molecular ultracold quantum simulators have emerged as an extremely innovative toolbox to study many-body physics which would otherwise be hard to probe in condensed matter systems or simulate numerically.
This is due to the high degree of precision that can be achieved in quantum optics experiments, which allows one to recreate prototypical models of quantum many-body systems under controlled settings~\cite{Zwerger:2003,Esslinger:2010,Bloch:2012,Ritsch:2013,Gross:2017,Blackmore:2018,Cooper:2019,Schaefer:2020,Mivehvar:2021}.
Among the numerous models that can be quantum simulated in ultracold systems, the Bose-Hubbard (BH) model is undoubtedly one of the most fundamental for describing strongly-correlated collective quantum behavior 
~\cite{Fisher_BH,Jaksch_cold,BH_experiment_1,BH_experiment_2}. 
Besides offering a prototypical realisation of the superfluid-Mott insulator phase transition~\cite{Fisher_BH,BH_experiment_1,Takafumi:2017,Lin:2019,Lin:2021}, the original BH model has been used as effective description of many other different physical systems, such as Josephson junction arrays~\cite{Fazio:2001, Bruder:2005}, granular and thin-film superconductors~\cite{Gersch:1963,Das:1999,Yurkevich:2001}, magnetic insulators~\cite{Giamarchi:2008,Zapf:2014}, and of course optical lattices~\cite{Jaksch:2005}.

Whereas the quantum simulation of the lowest-band BH model has been demonstrated in many different scenarios, its quantitative validity can become reduced by interactions~\cite{Oosten:2003,Larson:2009,Dutta:2011,Mering:2011,Luehmann:2012,Martikainen:2012,Lacki:2013,Xu:2016,Li:2022}.
This is particularly true for long-ranged interactions that appear in dipolar systems.
Dipolar Bose-Hubbard (DBH) models have been implemented using magnetic atoms ~\cite{EBH_magnetic_atoms} and may be implemented using the more strongly interacting polar molecules in the near future \cite{molony:2014,reichsoellner:2017,moses:2017}. 
These models have drawn interest due to the rich physics of strong off-site interactions that lead to phases such as density waves, supersolids, and the Haldane insulator~\cite{pollet:2010,capogrosso:2010,dalla:2006}.

Here, we quantitatively study the validity of the DBH model for a one-dimensional dipolar quantum simulator.
We contrast the physics of the underlying continuum-space description with the one of the lattice model it is supposed to recreate.
Across the two descriptions, we compare observables such as energies and density distributions. 
We also obtain a quantitative measure of their compatibility by calculating direct overlaps between their full many-body wavefunctions.
Our results show that for stronger DDI  (several times the recoil energy) and higher densities (above 0.67 particles per site) the single-band DBH description fails to correctly reproduce the physics of the continuum system.
Considering an effective two-band lattice description reduces this discrepancy, but still does not fully account for density distortions present in the continuum description.
Our work demonstrates that strong DDI require careful considerations when benchmarking the validity of dipolar quantum simulators and offers a method for quantitatively verifying the parameter regimes in which quantum simulation remains accurate.


\section{Physical Scenario}
\label{sec_physical_scenario}
We investigate the ground state of $N = 4$, $5$, or $6$ dipolar-interacting bosons of mass $m$ in a 1D optical lattice with $L = 9$ sites and infinite outer potential walls.
The potential between the walls is modelled in as
\begin{equation}
A(x) = \frac{A_0}{E_R} \sin^2 \left( \pi \frac{x}{L_0} \right),
\end{equation}
where $L_0 = \lambda/2$ is our chosen unit of length (the distance between two neighboring lattice sites).
We consider potential depths $A_0$ from $5 E_R$ to $10 E_R$, where $E_R=\frac{h^2}{2m \lambda^2}$ is the recoil energy, which acts as our unit of energy.
This optical lattice can be implemented straightforwardly in experiments by counter-propagating laser beams, whereas
the hard-wall boundaries can be engineered via a flat-bottom trap~\cite{Gaunt:2013,Navon:2021}. 
A combination of both has been employed in state-of-the-art experiments~\cite{Mazurenko:2017,Gall:2021}.
We will study filling fractions below one particle per site to investigate the effect of inter-site dipolar interactions on the validity of the lowest-band dipolar BH model (1BDBH) and the lowest-2-bands dipolar BH model (2BDBH).

We assume a strong transverse harmonic confinement, which regularises the short-range divergence of the DDI~\cite{continuum_1D_crystal_2, DDI_1D_regularisation}. 
To include this regularisation, we use the DDI potential
\begin{equation}
\mathcal{U}_V(x-x') = \frac{V L_0^3}{E_R (|x-x'|^3 + \alpha)},
\end{equation}
where $x$ and $x'$ are the particles' coordinates and $\alpha = 0.05$. 
$V$ thus denotes the (non-regularised) DDI strength between two bosons separated by one lattice site.
To reduce the population of multiply-occupied sites within the lowest band to $<1\%$, we consider a very strong contact interaction represented as a narrow Gaussian in the continuum calculations~\footnote{We remark that this term becomes irrelevant as the repulsive DDI performs this role more strongly for $V \geq 1.0 E_R$.},
%
\begin{equation}
\mathcal{U}_G(x,x') = \frac{V_G}{E_R \sqrt{2 \pi \sigma^2}}e^{-\frac{(x-x')^2}{2\sigma^2 L_0^2 }},
\end{equation}
where $V_G = 2.5 E_R$, and $\sigma = 0.05$.

\section{Lattice methods}
\label{sec_lattice_methods}
For a sufficiently deep optical lattice potential, the continuum system should map onto a DBH model through a tight-binding approximation. 
While the use of the DBH is widespread~\cite{non_standard_Hubbard,Biedron:2018,Hughes:2022,Lagoin:2022,Tamura:2022}, we briefly describe the derivation here due to its importance.

\begin{figure*}[!]
\includegraphics[width=\textwidth]{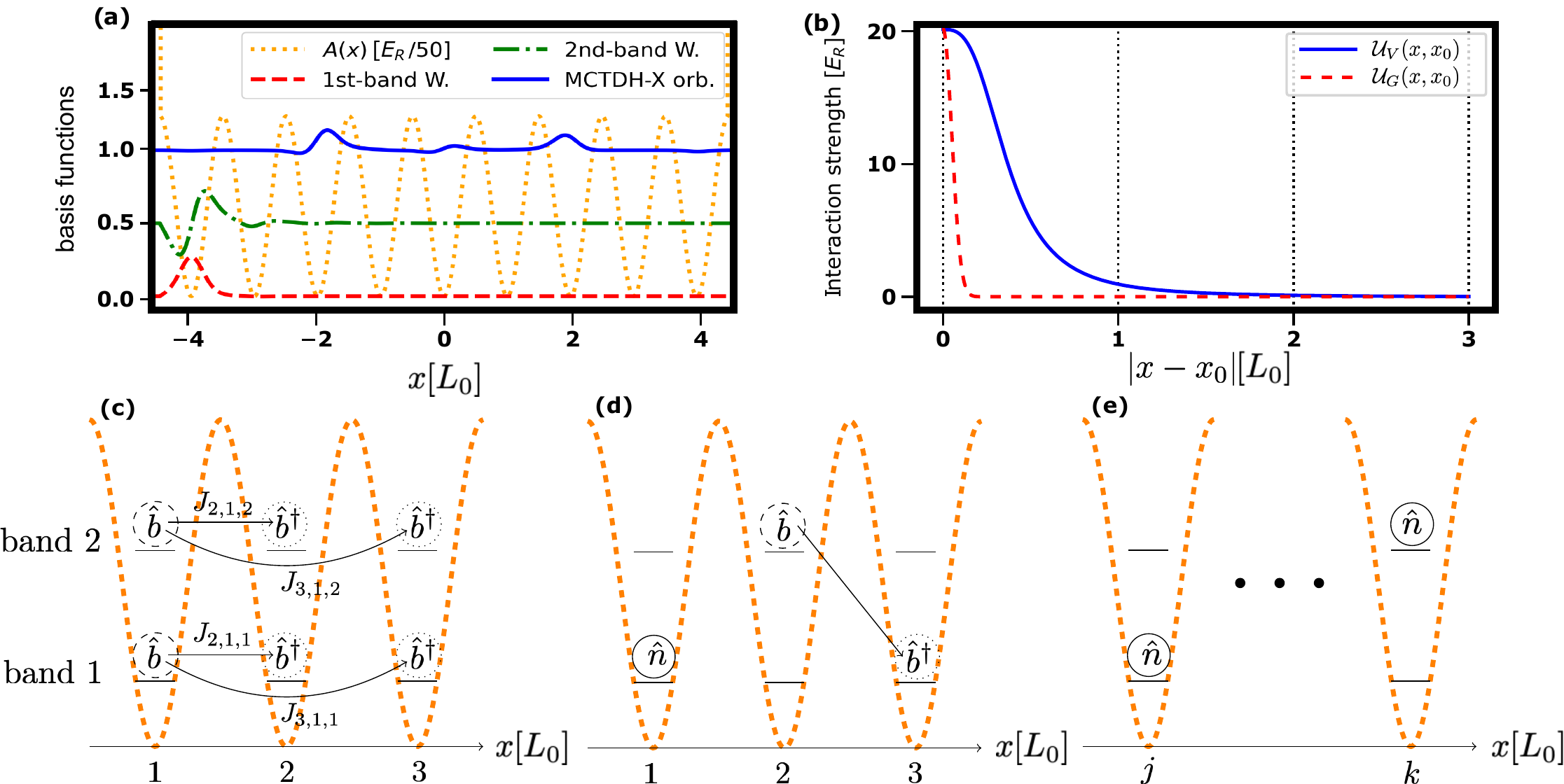}
\caption{(a) Examples of basis functions in the lattice description (1-band and 2-band finite-size Wannier functions) and in the continuum (a representative of the $M$ orthonormal MCTDH-X orbitals).
The functions are vertically shifted from 0 by multiple of $0.5$ to facilitate visualization.
(b) Spatial extent of interactions ($V=1.0 E_R$) for a particle centered at $x_0=0$ and another at $x$. The other panels show schematic illustration of some of the terms appearing in the 2BDBH model: (c) tunnelling, (d) density-induced tunnelling and (e) density-density interactions.}
\label{fig:diagram}
\end{figure*}

The single-particle Hamiltonian is given by the sum of kinetic and lattice potential terms
\begin{equation}
\mathcal{H} = \int dx ~ \hat{\Psi}^{\dagger}(x) \left[ -\frac{\hbar^2}{2m}\nabla^2 + A(x) \right] \hat{\Psi}(x), \label{eq_continuumH}
\end{equation}
where $\hat{\Psi}(x)$ annihilates a boson at position $x$ while obeying bosonic commutation relations and $m$ is the mass of a boson. 
In an infinite periodic lattice potential, the eigenvalues of this Hamiltonian form low-energy bands, labelled in ascending energy order by $\sigma$. 
While the eigenstates themselves are Bloch functions extended over the whole lattice, it is useful to describe the physics of the inter-particle interacting lattice model using Wannier functions $w_{j,\sigma}(x)$, which are superpositions of the Bloch functions of band $\sigma$ that become localised at lattice sites $j$ \cite{Wannier_original,Kohn_Wannier}.
We have used Wannier functions derived for finite-size lattices, which are not translationally-invariant (see appendix \ref{sec_Wannier_functions}).
Examples of such Wannier functions are shown in Fig.~\ref{fig:diagram}(a).

The Wannier functions provide a useful basis to decompose the boson field operator $\hat{\Psi}(x)$ using the equation
\begin{equation}
\hat{\Psi}(x) = \sum_{j,\sigma} w_{j,\sigma}(x) \hat{b}_{j,\sigma}, \label{eq_fieldop}
\end{equation}
where $\hat{b}_{j,\sigma}$ now annihilates a boson which has a spatial wavefunction given by $w_{j,\sigma}(x)$. 
If the lattice potential depth is very large compared with other energy scales, such as temperature and interactions, which could cause excitation to higher bands, the use of the lowest band is typically sufficient.
This might not be the case when interactions are strong and long-ranged, as for the DDI regimes considered in this work.
Because of that, we have performed separate calculations using the lowest band and the lowest two bands. 

When expanded in the basis of Wannier states, the single-particle continuum Hamiltonian in equation \eqref{eq_continuumH} creates `tunnelling' terms which cause particles to hop between sites within the same band. 
The tunnelling element $J_{j,k,\sigma}$ between site $j$ and site $k$ of band $\sigma$ is calculated using the integral
\begin{equation}
J_{j,k,\sigma} = - \int \mathrm{d}x \: w^*_{j,\sigma}(x) 
\left[
\frac{-\hbar^2}{2m}\nabla^2 + A(x)
\right]
w_{k,\sigma}(x). \label{eq_J}
\end{equation} 
These elements result in the single-particle tunnelling part of the BH lattice Hamiltonian
\begin{equation}
H_J = \sum_{j,k,\sigma} -J_{j,k,\sigma} \hat{b}^{\dagger}_{j,\sigma} \hat{b}_{k,\sigma} + H.c. \label{eq_HJ} 
\end{equation}
%
For the lattice potential depths studied, the tunnelling elements decay quickly with the distance between the sites.
Therefore, we neglect tunnelling beyond next-nearest neighbouring sites. 
The term for $j = k$ corresponds to a chemical potential, which costs energy for occupation of excited bands.

The lattice Hamiltonians corresponding to the inter-particle DDI and contact interaction are also calculated by integrating their elements in the Wannier basis~\footnote{We do not use the common approximation that these DDI decay as the inverse-cube of the distance between the site minima to improve the comparison between lattice and continuum results \cite{DDI_not_inverse_cube}.} 
For example, the lattice Hamiltonian for the DDI is calculated using
\begin{multline}
V_{i,\sigma_1,j,\sigma_2,k,\sigma_3,l,\sigma_4} = \frac{1}{2} \int dx \int dx' w_{i,\sigma_1}^*(x) w_{j,\sigma_2}^*(x') \\ \mathcal{U}_V(x-x') w_{k,\sigma_3}(x') w_{l,\sigma_4}(x),
\label{eq:V-matrix-el}
\end{multline}
where $V_{i,\sigma_1,j,\sigma_2,k,\sigma_3,l,\sigma_4}$ is the matrix element corresponding to the operator $\hat{b}^{\dagger}_{i,\sigma_1} \hat{b}^{\dagger}_{j,\sigma_2} \hat{b}_{k,\sigma_3} \hat{b}_{l,\sigma_4}$.
A sketch of some of the DBH terms appearing in the lattice Hamiltonian is given in Fig.~\ref{fig:diagram}(c) to (e).

\begin{figure}[!]
\includegraphics[width=\columnwidth]{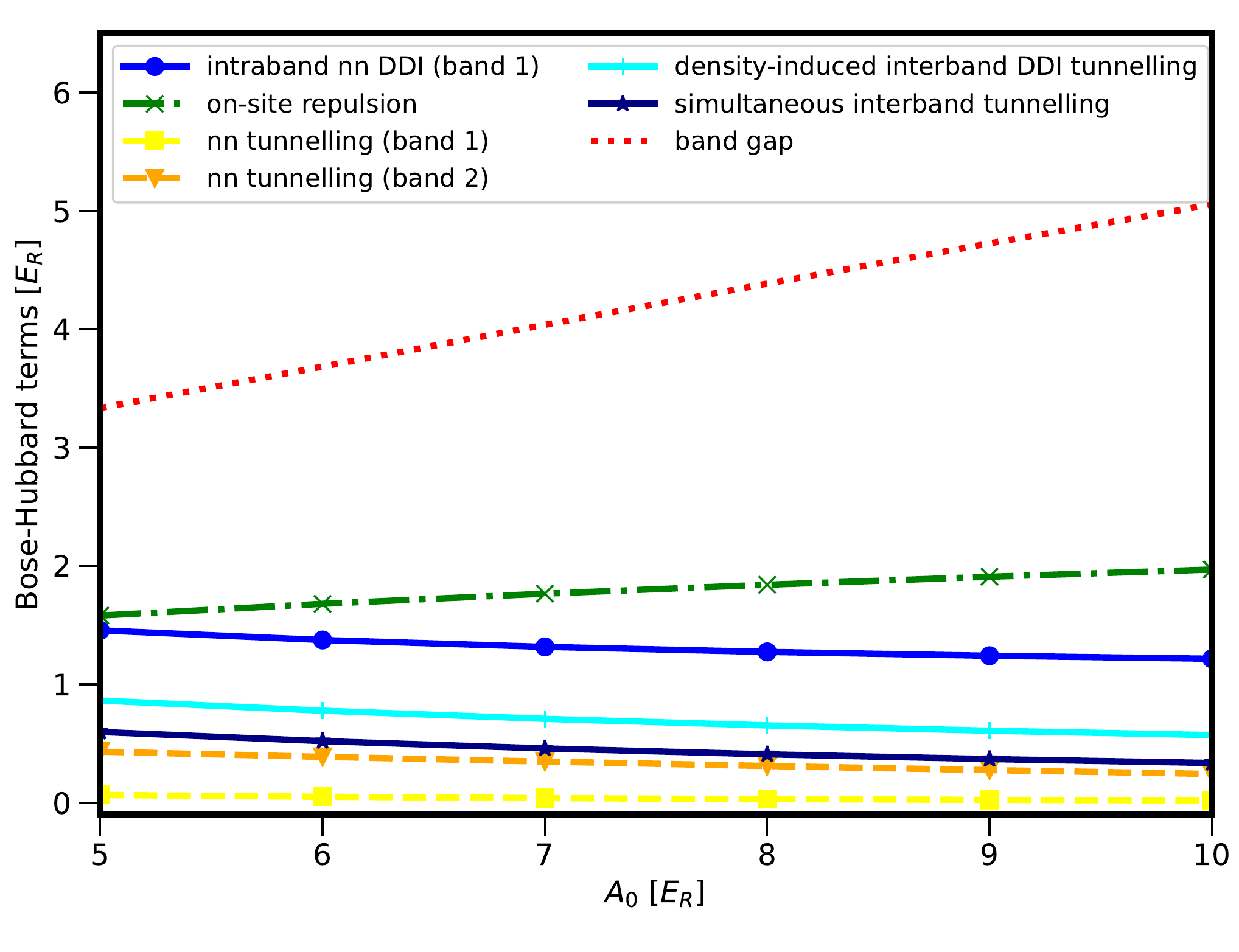}
\caption{Behavior of some nearest-neighbor (nn) DBH terms at the center-most sites of the lattice as a function of lattice depth $A_0$ and in comparison with the bandgap shown for $V = V_G = 1 E_R$. Solid lines: Selected DDI terms. Intraband nn DDI (band 1) corresponds to $V_{4,1,5,1,5,1,4,1}$   cf. Eq.~\eqref{eq:V-matrix-el}). Density-induced DDI tunnelling corresponds to  $|V_{4,1,5,1,5,2,4,1}|$. Simultaneous interband tunnelling corresponds to $|V_{4,2,5,1,5,2,4,2}|$. Dashed lines: nn tunnelling terms. Dash-dotted line: on-site repulsion. Dotted line: band gap $|J_{5,5,2}-J_{5,5,1}|$.}
\label{fig:DBH-terms}
\end{figure}

The long-range nature of the DDI means these elements generally decay more slowly with distance than the single-particle tunnelling elements. 
In our calculations, we have included all DDI and contact interaction elements where $i$, $j$, $k$, and $l$ are contained within three consecutive sites. 
We have included all DDI terms where $i = l$ and $j = k$ because these terms do not rely on the spatial overlap of far-separated Wannier functions. 
In the single-band model, these terms correspond to density-density repulsion $\hat{n}_i \hat{n}_j$, while in the two-band model, there are extra terms where bosons `tunnel' between the two bands within a site.
A quantitative comparison between band gap and the magnitude of some DBH terms, including nearest-neighbor DDI, intraband tunnelling, and density-induced interband DDI tunnelling, is given in Fig.~\ref{fig:DBH-terms}\footnote{Note that there are tens of different DBH terms which we included in the calculations, but to make the figure readable, we have only shown the most relevant ones.}

We solve the lattice model using exact diagonalization (ED) \cite{ED_BH} implemented in the QuSpin library \cite{QuSpin_part_1, QuSpin_part_2}. 
As the repulsive DDI and contact interaction strongly discourage multiple occupation of sites, we limited the number of bosons in each site of each band to $2$. 
The resulting wavefunction is denoted by $\ket{\psila}$ for the 1BDBH and by $\ket{\psilb}$ for the 2BDBH.

\section{Continuum Methods}
\label{sec_continuum_methods}
To benchmark the validity of the quantum simulator in implementing the DBH model, we also solve the many-body Schr\"{o}dinger equation directly for the continuum system 
which consists of the single-particle Hamiltonian of Eq.~\eqref{eq_continuumH} and the dipolar and contact interactions given by
\begin{align}
\mathcal{H}_{\text{int}} &= \int \mathrm{d}x \: \hat{\Psi}^{\dagger}(x) \hat{\Psi}^{\dagger}(x') 
\mathcal{U}_V(x,x') \hat{\Psi}(x') \hat{\Psi}(x) \nonumber \\
& \: + \int \mathrm{d}x \: \hat{\Psi}^{\dagger}(x) \hat{\Psi}^{\dagger}(x') 
\mathcal{U}_G(x,x') \hat{\Psi}(x') \hat{\Psi}(x).
\end{align}
%

To calculate the continuum ground state $\ket{\psim}$, we employ the MultiConfigurational Time-Dependent Hartree method for bosons (MCTDH-B)~\cite{Streltsov:2006, Streltsov:2007, Alon:2007, Alon:2008}, implemented by the MCTDH-X software~\cite{Lode:2016,Fasshauer:2016,Lin:2020,Lode:2020,MCTDHX}.
MCTDH-X implements a variational optimization procedure by decomposing the many-body wavefunction with an adaptive basis set of $M$ time-dependent single-particle orbitals. 
The orbitals form a very different basis set than the Wannier function, and they can be non-local (see Fig.~\ref{fig:diagram}(a)).
In our calculations with $L=9$ sites, we employ $M=9$ for simulations involving $N=4,5$ particles and $M=18$ for simulations involving $N=6$ particles.
This choice of $M$ thus allows us to faithfully describe each site with respectively one ($M=9$) or two ($M=18$) orbitals, and to project the corresponding wave function to the single-band or double-band lattice basis.

\begin{figure}[!]
\includegraphics[width=\columnwidth]{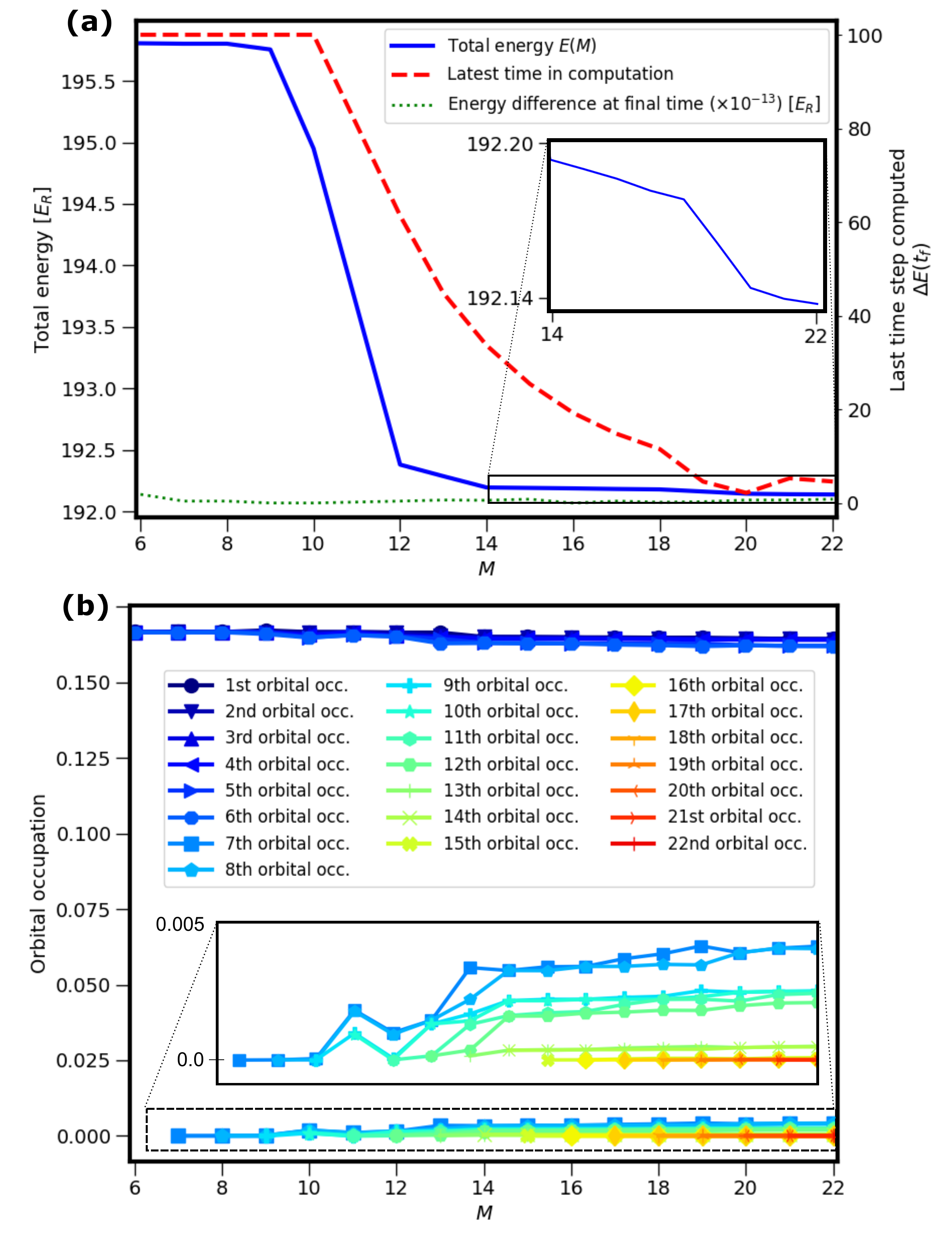}
\caption{
(a) Energy convergence in continuum calculations for $N=6$, $A_0=10 E_R$, and $V=10.1 E_R$. 
The solid blue line is the change of ground-state energy as a function of $M$. 
The dotted green line represents the difference in energy at the end of each relaxation with different $M$.
The red dashed line indicates the computation time step needed to achieve the given energy convergence.
(b) Orbital occupation convergence in continuum calculations for the same parameters. The occupations converge to a fixed value as $M$ is increased, as evinced also by the negligible occupation of the additional orbitals for $M \ge 15$ (inset).}
\label{fig:M-dependence}
\end{figure}

We have verified that the contribution of higher orbitals is negligible both in terms of occupation and of change in ground-state energy, as shown in Fig.~\ref{fig:M-dependence}.
For instance, for $N=6$ we find that already beyond $M=15$, the change in ground-state energy that occurs when adding more orbitals is less than $0.004 E_R$ and their occupation is below $0.00001$.
As we shall see when comparing continuum calculations with lattice ones, this change in energy is three orders of magnitude smaller than the discrepancy observed between continuum and 1BDBH lattice model.

\section{Method Comparison}
\label{sec_comparison}
We use several quantities to compare the results of the lattice and continuum methods. 
The first is the total kinetic, potential, and interaction ground state energy. 
We then compare the boson density expectation values on the fine spatial grid used by the MCTDH-X calculations, where $\langle \psil|\hat{\Psi}^{\dagger}(x)\hat{\Psi}(x)| \psil \rangle$ is calculated using equation \eqref{eq_fieldop}. 
Finally, we project the MCTDH-X wavefunction $\ket{\psim}$ to the Hilbert space of the lowest band to obtain $\ket{\psimtola}$ or of the lowest two bands to obtain $\ket{\psimtolb}$.
For details, see appendix~\ref{sec_overlap}.
We evaluate the projection magnitude onto these Hilbert spaces as $P_1 = \langle \psimtola |\psimtola \rangle$ and $P_2 = \langle \psimtolb | \psimtolb \rangle$, where $P_2 \geq P_1$. 
We then calculate the overlap of these projections with the ED wavefunctions as $f_1 = |\langle \psila| \psimtola \rangle|^2$ and $f_2 = |\langle \psilb| \psimtolb \rangle|^2$, where $f_1 \leq P_1$ and $f_2 \leq P_2$.
This gives a \textit{quantitative} measure to compare the accuracy of the lattice representation in the continuum system.

\section{Results}
\label{sec_results}
We now present the results obtained by comparing the observables mentioned above across continuum and lattice descriptions.
We will first examine lower densities ($N=4,5$ particles), where the 1BDBH model is well reproduced (even when the filling fraction is incommensurate with the density wave order), and then focus on higher densities ($N=6$), where higher bands must be considered.

\subsection{Lower densities -- $N=4,5$}
We begin by considering the case of $N = 5$ bosons. 
The filling fraction here is chosen such that the ground state is a superfluid at weak DDI and a density wave (DW) $\ket{1 0 1 0 1 0 1 0 1}$ with a large energy gap at strong DDI. 
We compare this to the case of $N = 4$, which has a lower filling fraction that is not commensurate with the density wave order. 
This leads to a much smaller gap between the lowest energy states of the lowest-band dipolar BH model at strong DDI. 

\begin{figure}[!]
\includegraphics[width=\columnwidth]{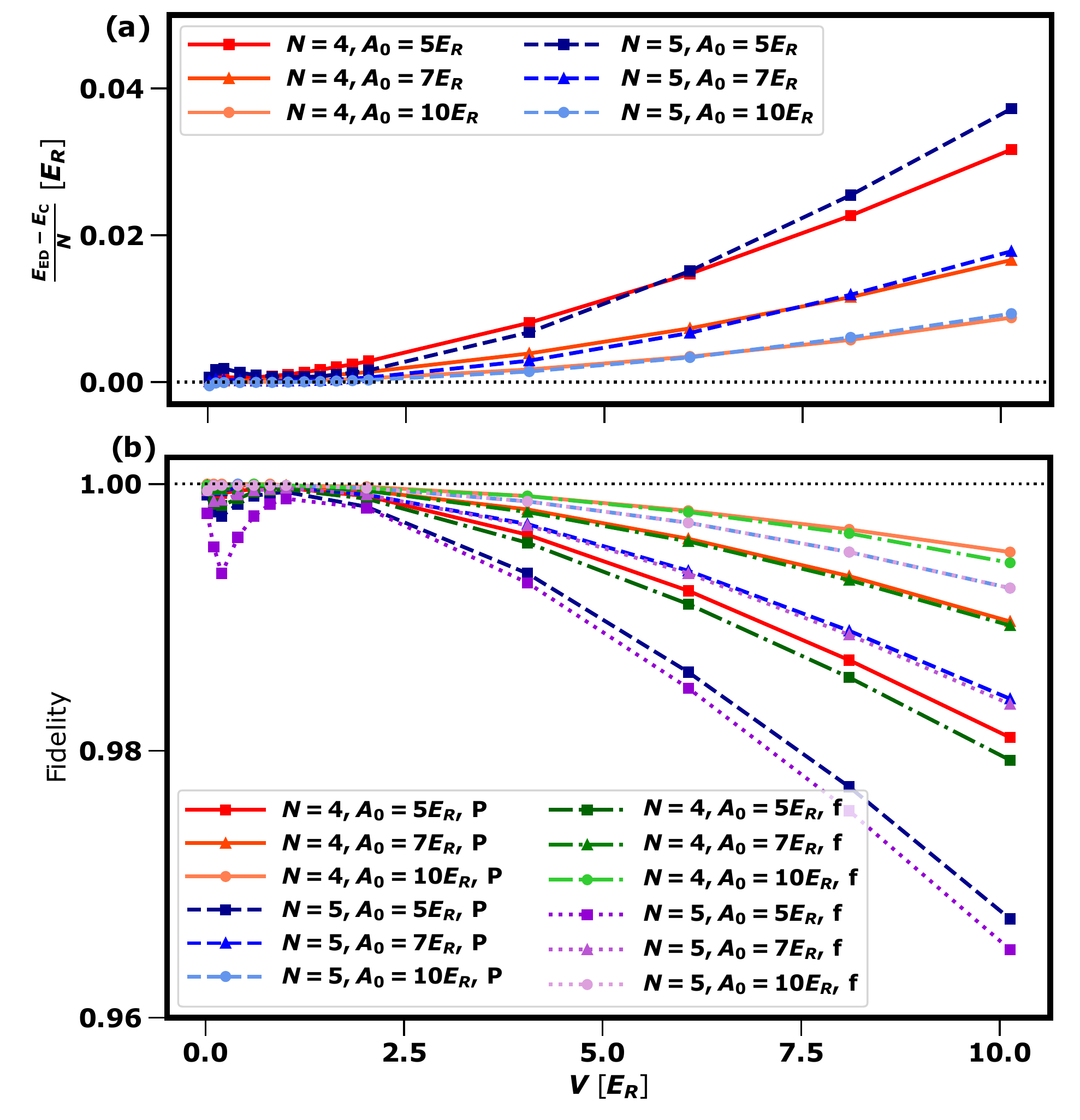}
\caption{
Comparison between lattice and continuum results for $N = 4$ and $N = 5$. 
(a) Comparison of ground state energies per particle for lattice ($E_{ED}$) and continuum ($E_C$). 
(b) Projection magnitude $P$ of $\ket{\psim}$ onto lowest band and overlap $f$ with corresponding ED ground state.}
\label{fig:energy-fid-N4-N5}
\end{figure}

\begin{figure}[!]
\includegraphics[width=\columnwidth]{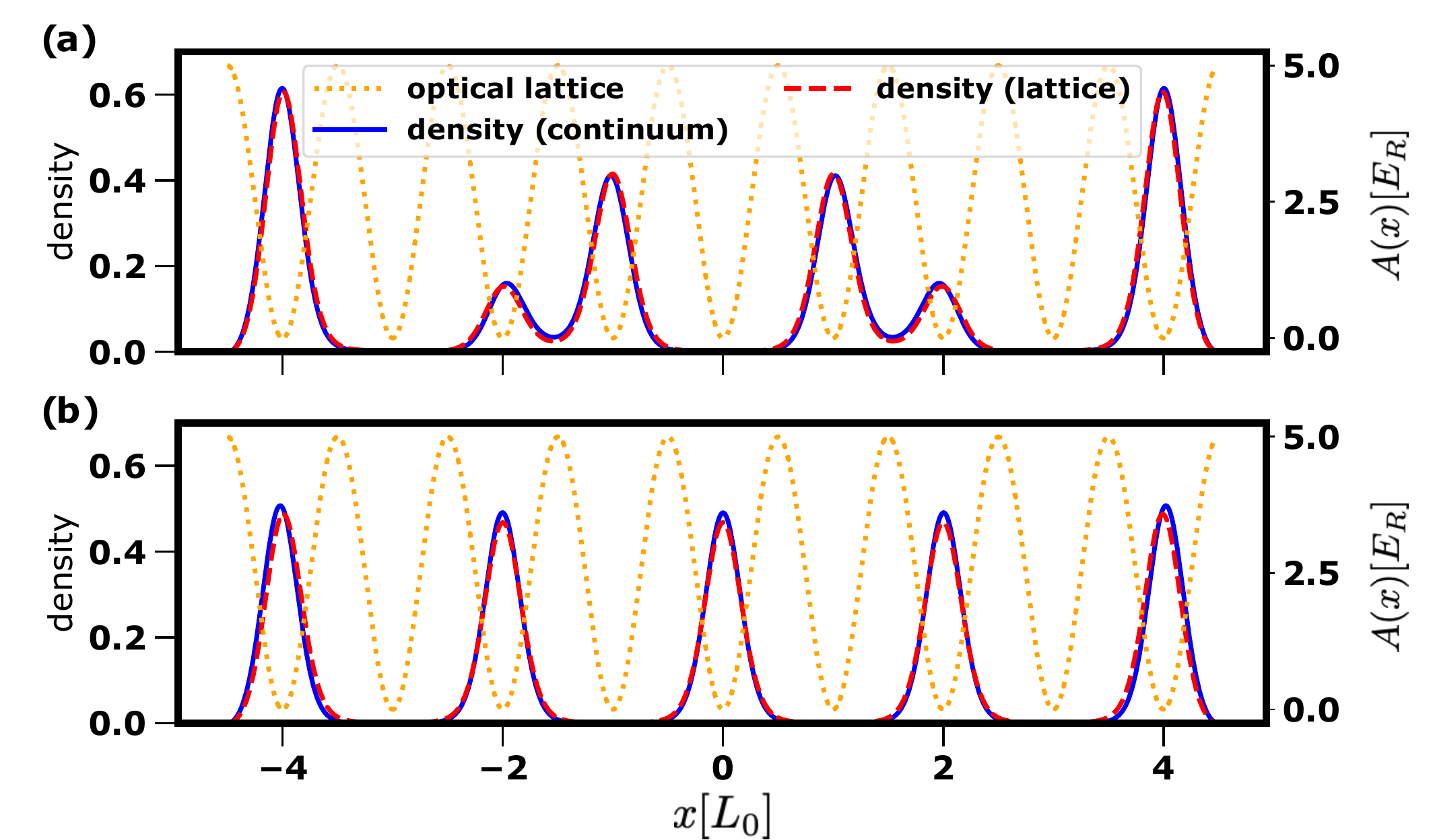}
\caption{
Comparison of densities between methods for (a) $N = 4$, $A_0 = 5 E_R$, $V = 10 E_R$ and 
(b) $N = 5$, $A_0 = 5 E_R$, $V = 10 E_R$.
The density is normalised to integrate to 1.
}
\label{fig:dens-fid-N4-N5}
\end{figure}

Figure \ref{fig:energy-fid-N4-N5}(a) compares the ground state energies in the two methods. 
For small DDI, the good quantitative agreement for $A_0 = 10 E_R$ reduces slightly for weak lattice potentials where the lattice description is less appropriate.
There are larger discrepancies between methods for strong DDI, 
an effect which increases with shallower potentials and higher particle density as expected. 
For $N = 5$, $A = 5 E_R$, $V = 10.1 E_R$, the lattice energy is $\approx 0.2 E_R$ ($\approx 1\%$) larger than the continuum energy. 

Figure \ref{fig:energy-fid-N4-N5}(b) shows the projection magnitude of the MCTDH-X ground state onto the lowest band and onto the corresponding ED ground state itself. 
This shows quantitatively that the energy discrepancy at strong DDI coincides with a small population of higher bands. 
For $A_0 = 5 E_R$ and $N = 5$, there is a small reduction in $P_1$ and an increase in the energy gap between methods for $V \approx 0.2 E_R$ where the larger tunnelling at the weak lattice potential means there is still a small occupation in the nominally-empty sites of the DW. 
This effect reduces as the DW becomes fully established and the particles are separated further. 

Figures \ref{fig:dens-fid-N4-N5}(a) and (b) show the continuum densities for $A = 5 E_R$, $V = 10.1 E_R$ for $N = 4$ and $N = 5$ respectively. 
The densities of the two methods are visually similar except for small warping of the continuum on-site density to minimize dipolar repulsion. 
For $N = 4$, this slightly displaces the density on the edge sites  further out and the density on the adjacent sites towards the centre of the lattice. 
For $N = 5$, since the filling is commensurate with the density wave, there is less benefit in displacing the on-site density except pushing the edge sites further outwards and narrowing the peaks.

\subsection{Higher densities -- $N=6$}

We now present results for $N = 6$, where the quantum simulation of the single-band dipolar BH model in the continuum system becomes drastically less precise because of increased DDI effects.
Like $N = 4$ the energy gaps to low-lying excited states of the lattice model are small due to incommensurate filling and the lattice ground state is significantly entangled for the range of $V$ we cover.  
We empirically find that the continuum results for $N = 6$ are not converged with respect to $M = 9$ single-particle orbitals for $V \geq 0.5 E_R$ and instead require $M = 18$.
We therefore compare these results with both the lowest-band dipolar BH model and the corresponding two-band model, which features the same number of single-particle orbitals as the continuum calculations.

\begin{figure}[!]
\includegraphics[width=\columnwidth]{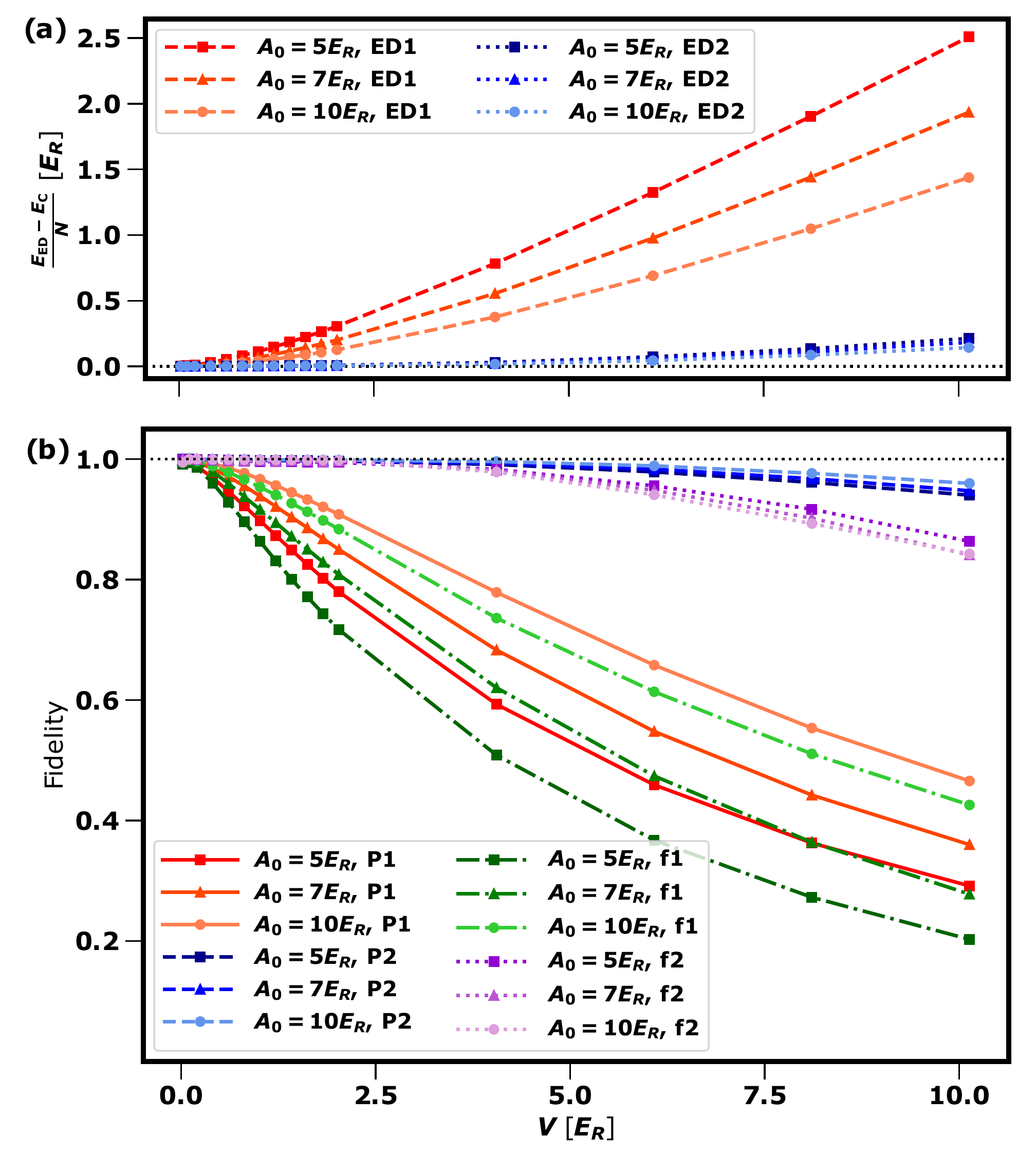}
\caption{
Comparison between lattice and continuum results for $N = 6$. 
(a) Difference in total energy between continuum results and ED with one and two bands. 
(b) Projection magnitudes of $\ket{\psim}$ onto lowest band (P1) and lowest two bands (P2) and overlaps with corresponding ED ground states (f1/f2).}
\label{fig:N6-energy-fid}
\end{figure}

\begin{figure}[!]
\includegraphics[width=\columnwidth]{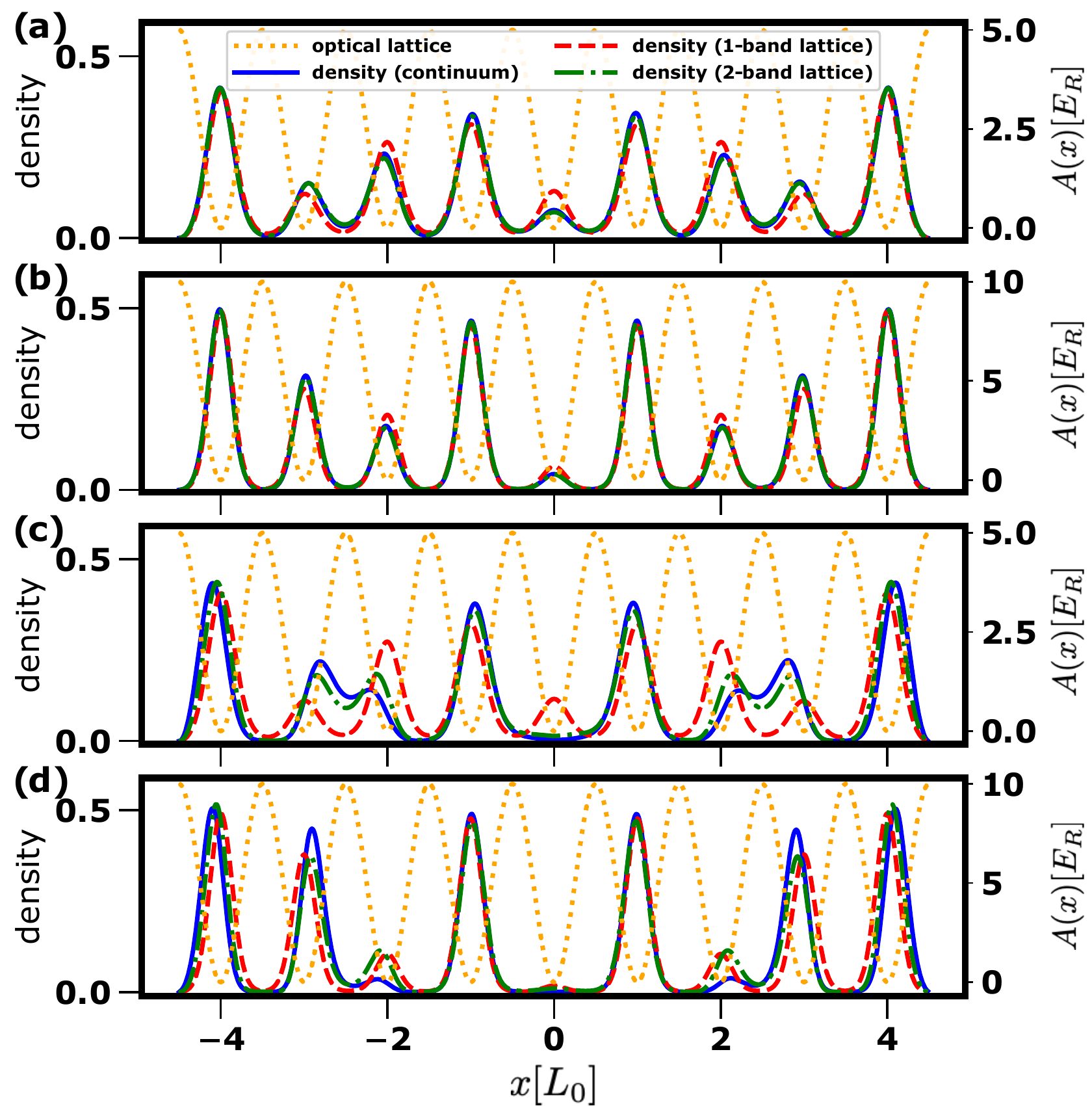}
\caption{
Comparison of densities between methods for $N = 6$, using the common legend in (a). 
(a) shows $A_0 = 5 E_R$, $V = 1.01 E_R$, 
(b) shows $A_0 = 10 E_R$, $V = 1.01 E_R$, 
(c) shows $A_0 = 5 E_R$, $V = 10.1 E_R$, 
and (d) shows $A_0 = 10 E_R$, $V = 10.1 E_R$.
The density is normalised to integrate to 1.
}
\label{fig:N6-dens}
\end{figure}

In Fig.~\ref{fig:N6-energy-fid}(a), the energies of the continuum calculations, single-band ED, and two-band ED agree well at weak DDI as expected. 
For strong DDI, however, the continuum calculations are able to achieve a significantly lower energy than the single-band model and slightly lower than the two-band model. 
In Fig.~\ref{fig:N6-energy-fid}(b), $P_1$ (and therefore $f_1$) decreases below $0.5$ for all potential depths studied, quantitatively showing the strong population of excited bands for strong DDI. 
As expected, $P_1$ and $P_2$ decrease with lattice potential depth, although this pattern is not strictly observed with $f_1$ and $f_2$ due to the close energy competition of states within and outside the specified bands.

For $N=6$ there is a clearer repositioning of particle density within sites to reduce the DDI repulsion [Fig.~\ref{fig:N6-dens}(a)-(d)], because the greater average density leads to at least one pair of neighbouring sites with particles.
While the 1BDBH (red dash lines) fails at correctly recreating this effect, the 2BDBH is at least able to qualitatively capture it because superpositions of occupation of the lowest two bands are offset from the lattice minimum.

\section{Discussion}
\label{sec_discussion}
The much greater deviations between the lattice and continuum methods for $N = 6$ compared to $N = 4$ and $N = 5$ suggest that the quantitative validity of the 1BDBH model at strong DDI is strongly dependent on the filling pattern of the ground state. 
The occupations of nearby sites significantly affect the warping of the on-site density. 
This would complicate the application of on-site occupation-dependent Wannier functions and BH parameters, which have been used for strong short-range interactions \cite{broaden_Wannier_2, broaden_Wannier_2, broaden_Wannier_3}. 
This effect may be more important for more complicated particle arrangements in higher dimensions or for attractive interactions where the particles preferentially occupy nearby sites.
For higher filling fractions above one particle per site, more dramatic dipolar crystal states with multiple density peaks per lattice site have been predicted, clearly populating excited bands~\cite{chatterjee:2018,chatterjee:2019,chatterjee:2020}.

Experimentally, BH models showing dynamical effects of the DDI have been implemented using magnetic atoms \cite{EBH_magnetic_atoms}. 
In that example, the DDI between the two atoms at a distance of one lattice site was around $1\%$ of the recoil energy, deep within the regime of quantitative agreement between the lowest-band dipolar BH model and the continuum methods for moderate lattice potentials.
For the most strongly dipolar polar molecules, this is not necessarily the case. 
Taking the example of NaCs, which has a large electric dipole moment of up to $4.7$D ($2.6$D of which has been implemented in recent experiments \cite{NaCs_massive_dipole_moment}), the repulsive DDI between two molecules polarised perpendicularly to the lattice at a distance of $532$nm could be up to $\approx 20$kHz. 
This is over $15 \times$ the recoil energy $E_R$ in a lattice of wavelength $\lambda = 1064$nm, suggesting that the DDI strengths required to compete with moderate lattice potentials may be realizable in near-future experiments. 
We considered filling fractions of around one boson per two sites, which is somewhat larger than previously observed filling fractions of $30\%$ for bosonic polar molecules in optical lattices \cite{polar_molecules_lattice_4}. 

Before presenting our concluding remarks, a comment about the energy scale is in order.
When the DDI are very strong and repulsive and the density is below one particle per site, the particles rearrange themselves to reduce their average interaction energy well below the value of $\mathcal{U}_V \approx 0.952 V $ at one lattice site distance.
For example, the two-band lattice results for $N=6$ and $A_0=10 E_R$ have DDI energy per particle of $3.34E_R$ when $V = 10.1 E_R$. 
For the single-band results for $N = 5$ particles, the DDI energy per particle at $V = 10.1 E_R$ is $1.19 E_R$ because the particles are able to arrange in the density wave.
Nevertheless, these interaction energies are still quite high, and the outer box-like potential wall introduced in section II is  required to stop the particles from trying to escape the lattice.
Furthermore, the construction of the lattice model on the non-interacting single-particle band structure is a rough approximation to the physics when the DDI is comparable to the lattice potential.
Our comparison with the continuum calculations shows however that the 1BDBH model, despite being inaccurate, is still well-behaved.

\section{Conclusions}
\label{sec_conclsions}
We quantitatively studied the breakdown of the lowest Hubbard band description in 1D repulsively-interacting systems, finding that the on-site wavefunctions are distinctly warped by the presence of particles on other sites.
We performed a direct comparison between continuum and lattice models, not only by calculating energy and density, but also by constructing many-body wavefunction overlaps directly between the two methods.
The message of our results is twofold.
On the one hand, they highlight that the DDI strongly influences the parameter regimes where dipolar quantum simulators correctly reproduce single-band BH models.
On the other hand, they show that experimentally-feasible parameters can access 
the regimes in which the DDI significantly populates higher bands for finite-size lattices.
More generally, our method of measuring the overlap between lattice and continuum many-body wavefunctions provides a robust and quantitative way of verifying the compatibility between the two descriptions.
This should help determine the correct parameter regimes of operation for near-term dipolar quantum simulators. 

Future work could include a more general quantification of the validity of lowest-band dipolar BH models in different geometries, including periodic boundary conditions, higher dimensions, and attractive interactions.
Our study could also be extended to focus on time evolution, where significant discrepancies between lowest-band lattice and continuum methods were found for contact interactions~\cite{Sakmann_BH_bad,major:2014}. 
The breakdown of the lowest-band BH model may have more significant consequences for the physics at weaker lattice potential or anisotropic DDI, where interactions may encourage particles to occupy neighbouring sites, or in cases where the DDI introduces more intricate competition of interactions within the BH model itself.

\acknowledgements
The authors would like to acknowledge the use of the University of Oxford Advance Research Computing (ARC) facility in carrying out this work~\cite{ARC}. 
We also acknowledge computation time on the ETH Zurich Euler cluster and at the High-Performance Computing Center Stuttgart (HLRS).
We thank Ofir Alon, Hongmin Gao, and Jordi Mur-Petit for helpful discussions.
We thank Rui Lin, Marcos Rigol, and Jakub Zakrzewski for useful comments on the manuscript.
This work is supported by the Cluster of Excellence `CUI: Advanced Imaging of Matter' of the Deutsche Forschungsgemeinschaft (DFG) - EXC 2056 - project ID 390715994, by the U.K. Engineering and Physical Sciences Research Council (EPSRC) Grants no. EP/P01058X/1 (QSUM) and no. EP/P009565/1 (DesOEQ), by a Simons Investigator Award, and by the Austrian Science Foundation (FWF) under grant P-32033-N32.
\newpage


\appendix

\section{Wannier functions}
\label{sec_Wannier_functions}

This appendix shows the derivation of the Wannier functions for finite lattices. 
It is based on that in the supplemental material of Ref \cite{cDMRG} and is functionally equivalent except for a small extension for the second band. 

We consider a lattice potential $A = A_0 \cos^2(\pi x)$ defined between $x = 0$ and $x = L$ with hard potential walls (after calculating the Wannier functions, we translate this to be symmetric about $x = 0$). The Wannier functions must vanish at the potential walls, so we build them using the basis functions $f_m(x) = \sqrt{2} \sin(m \pi \frac{x}{L})$ where $m$ is a positive integer. For the lowest bands of $L = 9$, $m \leq 800$ causes negligible truncation. The kinetic Hamiltonian $\hat{H}_K$ is diagonal in these basis functions, where each basis function has kinetic energy $(\frac{m}{L})^2 E_R$. The Hamiltonian for the lattice potential $H_P$ also couples basis states and is given by
%
%
\begin{align}
\hat{H}_P &= A_0 \sum_m \frac{1}{2}\ket{m}\bra{m} + \frac{1}{4}\ket{m}\bra{m+2L} \nonumber \\
& \qquad + \frac{1}{4} (\ket{m+2L}\bra{m} -\ket{m}\bra{-m+2L}),
\end{align}
where elements outside the chosen maximum value of $m$ are excluded. The single-particle Hamiltonian is the sum of the kinetic and potential terms. Due to its structure, this Hamiltonian only couples the basis functions in groups defined by $(m \pm q) \mathrm{mod} (2L) = 0$ labelled by $q$, where $q$ takes integer values from $0$ to $L$. Excluding small truncation effects, there are half the number of basis functions for $q = 0$ and $q = L$ each as for all other values of $q$. As in Ref \cite{cDMRG}, the lowest band is spanned by the lowest eigenstate for each value of $1 \leq q \leq L$. Meanwhile it is clear that the second-lowest eigenvalues for $1 \leq q \leq L-1$ are similar to the lowest eigenvalue for $q = 0$ and are far below the second-lowest eigenvalue for $q = L$. This motivates our identification of the lowest eigenstate for $q = 0$ and the second-lowest eigenstates for $1 \leq q \leq L-1$ as the second band.

As in Ref \cite{cDMRG}, the Wannier functions within the space of a band are found by taking the eigenstates of the position operator
\begin{equation}
\hat{X} = \sum_{m_1,m_2} ((-1)^{(m_1+m_2)}-1)\frac{4 L m_1 m_2}{\pi^2 (m_1^2-m_2^2)^2} \ket{m_1}\bra{m_2}
\end{equation}
These can be converted into functions in real space using the definition of the basis functions. The BH model parameters can be found by calculating the matrix elements of the kinetic Hamiltonian and any additional interaction Hamiltonians in the basis defined by the Wannier functions. For the DDI and contact interaction, we numerically integrated over real space for this purpose. The tunnelling elements and on-site energies for both the lowest and second bands away from the edges of the lattice are in good quantitative agreement with the equivalent elements for infinite lattices calculated using the methods in Ref \cite{MLGWS_paper}.

\section{MCTDH-X}
\label{sec_MCTDHX}

In this section we review the main theory behind the MultiConfigurational Time-Dependent Hartree method for indistinguishable particles implemented by the MCTDH-X software~\cite{Alon:2008,Lode:2016,Fasshauer:2016,Lin:2020,Lode:2020,MCTDHX}.
MCTDH-X is based on a variational optimization procedure in which the many-body wavefunction is decomposed into an adaptive basis set of $M$ time-dependent single-particle wavefunctions, called orbitals. 
Using the time-dependent variational principle in imaginary time, MCTDH-X optimizes both the coefficients and the orbitals to relax the system to its ground state.

The starting point of this approach is the total many-body Hamiltonian in second quantized formulation, which is composed of one- and two-body operators:
\begin{align} 
\hat{\mathcal{H}}&=\int dx \hat{\Psi}^\dagger(x) \left\{\frac{p^2}{2m}+A(x)\right\}\hat{\Psi}(x) \nonumber\\
&+\frac{1}{2}\int dx \hat{\Psi}^\dagger(x)\hat{\Psi}^\dagger(x')W(x,x')\hat{\Psi}(x)\hat{\Psi}(x').
\end{align}
The function $V(x)$ represents a one-body potential, while $W(x,x')$ encodes two-body interactions.
In the case of the present work, $A(x)$ is the optical lattice, while $W(x,x')$ is the sum of Gaussian repulsion and dipole-dipole interactions.

To time evolve the many-body Schr\"{o}dinger equation (either in real or imaginary time), the state of the system is first decomposed into $M$ orbitals -- through the following ansatz:
\begin{eqnarray}
	|\psim\rangle=\sum_{\mathbf{n}} C_\mathbf{n}(t)\prod^M_{k=1}\left[ \frac{(\hat{b}_k^\dagger(t))^{n_k}}{\sqrt{n_k!}}\right]|0\rangle. \label{eq:def_psim}
\end{eqnarray}
Here, $\mathbf{n}=(n_1,n_2,...,n_k)$ is the number of atoms in each orbital, which is subject to the global constraint $\sum_{k=1}^M n_k=N$, with $N$ the total number of particles.
Furthermore, in our notation $|0\rangle$ denotes the vacuum and $\hat{b}_i^\dagger(t)$ represents the time-dependent operator that creates one boson in the $i$-th working orbital $\psi_i(x)$, \textit{i.e.}:
\begin{eqnarray}
	\hat{b}_i^\dagger(t)&=&\int \mathrm{d}x \: \psi^*_i(x;t)\hat{\Psi}^\dagger(x;t) \:  \\
	\hat{\Psi}^\dagger(x;t)&=&\sum_{i=1}^M \hat{b}^\dagger_i(t)\psi_i(x;t). \label{eq:def_psi}
\end{eqnarray}
The equations of motion for the coefficients $C_\mathbf{n}(t)$ and the working orbitals $\psi_i(x;t)$ can be obtained by applying the time-dependent variational principle~\cite{TDVM81}, from which the real or imaginary time evolution of the system then follows.
The number of orbitals influences the accuracy of the algorithm: with a single orbital $M=1$, MCTDH-X is equivalent to a mean-field Gross-Pitaevskii description, while as $M \to \infty$ the method becomes numerically exact.

With the MCTDH-X method it is possible to compute $N$-body reduced density matrices from the working orbitals. 
For example, the one-body reduced density matrix can be calculated as
\begin{eqnarray}
\rho^{(1)}(x,x') = \sum_{kq=1}^M \rho_{kq}\psi_k(x)\psi_q(x' ),
\label{eq:red-dens-mat}
\end{eqnarray}
with
\begin{eqnarray}
\rho_{kq} = \begin{cases}
\sum_\mathbf{n} |C_\mathbf{n}|^2 n_k, \quad & k=q \\
\sum_\mathbf{n} C_\mathbf{n}^* C_{\mathbf{n}^k_q} \sqrt{n_k(n_q+1)}, \quad & k\neq q \\
\end{cases}.
\end{eqnarray}
Here, the sum enumerates all possible configurations of $\mathbf{n}$.
The notation $\mathbf{n}^k_q$ refers to the configuration where one atom is removed from orbital $q$ and then added to orbital $k$.
From the one-body reduced density matrix it is then straightforward to calculate the one-particle density as its diagonal elements:
\begin{equation}
	\rho(x) = \rho^{(1)}(x,x)/N
\end{equation}

\section{Construction of lattice/continuum overlaps}
\label{sec_overlap}

In this section, we provide details of the calculation of the projection of the continuum wavefunction into the lattice Hilbert space, and the overlap of the continuum and lattice wavefunctions. The structure of the continuum wavefunction is stated in equation \ref{eq:def_psim}. The lattice wavefunction has a similar form
\begin{eqnarray}
	|\psil\rangle=\sum_{\mathbf{n'}} C'_\mathbf{n'}\prod^{L}_{k'=1}\prod^{N'_b}_{\sigma'=1} \left[ \frac{(\hat{b'}_{k',\sigma'}^\dagger)^{n'_{k',\sigma'}}}{\sqrt{n'_{k',\sigma'}!}}\right]|0\rangle. \label{eq:def_psil}
\end{eqnarray}
where we have used primes to denote lattice quantities. $N'_b$ is the number of bands in the lattice model. $\mathbf{n'}=(n'_{1,1},...,n'_{k',\sigma'})$ is the number of atoms in each Wannier function, which is subject to the global constraint $\sum_{k'=1}^{L} \sum_{\sigma'=1}^{N_b} n'_{k',\sigma'}=N$, with $N$ the total number of particles. $\hat{b'}_{i',\sigma'}^\dagger$ represents the operator that creates one boson in the $i'$-th Wannier function of the $\sigma'$-th band $w_{i',\sigma'}(x)$.

To calculate the $\psimtol$ and $f$, we used the inner product of the single-particle basis functions for the two methods i.e. the working orbitals $\psi_i(x)$ for the continuum calculations and the Wannier functions $w_{i,\sigma}(x)$ for the lattice calculations. For this we calculated the integrals
\begin{equation}
O_{i,(i',\sigma')} = \int \mathrm{d}x  \: \psi_i(x)^* w_{i',\sigma'}(x) 
\label{eq:overlap_integral}   
\end{equation}

These overlaps can be used to project the annihilation operator for the MCTDH-X orbitals into the lattice Hilbert space using
\begin{equation}
\hat{b}_i = \sum_{i',\sigma'} O_{i,(i',\sigma')} \hat{b}_{i',\sigma'}, \label{eq:annihilation_correspondence}
\end{equation}
which in turn allows the projection of the MCTDH-X wavefunction onto the lattice Hilbert space $\ket{\psimtol}$ to be calculated using equation \ref{eq:def_psim}. The overlap with the ED wavefunction $f = |\langle \psil| \psimtol \rangle|^2$ can then be calculated easily in the lattice Hilbert space.


\bibliography{lattice-MCTDH-X_bib} 

\end{document}